\documentclass[12pt,a4paper]{article}
\usepackage{amsmath}
\usepackage{amssymb}
\usepackage{amsfonts}
\usepackage{amsthm}
\usepackage{bbm}
\usepackage{graphicx}
\usepackage{mathrsfs}
\usepackage{geometry} 
\usepackage{cases}
\usepackage{enumerate}

\theoremstyle{remark}

\newcommand{\rd}{\mathrm{d}}

\usepackage{indentfirst}
\usepackage{cite}
\usepackage{color}
\usepackage[hyperindex=true,
          pdfstartview=FitH,
          bookmarksnumbered=true,
          bookmarksopen=true,
          citecolor=blue,
          linkcolor=red,
          colorlinks=true, 
          unicode]{hyperref}
\begin{document}

\title{$f$(Ricci) Gravity}
\author{Chuanyi Wang
and Liu Zhao\thanks{Correspondence author.}\\
School of Physics, Nankai University, Tianjin 300071, China
\\
{\em email}: \href{mailto:2120170128@mail.nankai.edu.cn}
{2120170128@mail.nankai.edu.cn}
and
\href{mailto:lzhao@nankai.edu.cn}{lzhao@nankai.edu.cn}
}
\date{}
\maketitle

\begin{abstract}
$f$(Ricci) gravity is a special kind of higher curvature gravity whose bulk Lagrangian density is 
the trace of a matrix-valued function of the Ricci tensor. It is shown that, under some
mild constraints, $f$(Ricci) gravity admits Einstein manifolds as exact vacuum solutions, 
and can be ghost free around maximally symmetric Einstein vacua. 
It is also shown that the entropy for spherically symmetric black holes in $f$(Ricci) gravity
calculated via Wald's method and the boundary Noether charge approach proposed by Majhi
and Padmanabhan are in good agreement.
\end{abstract}

\section{Introduction}

Even after over a hundred years of intensive study, Einstein's general theory of 
relativity (GR) remains to be the most important theory of gravity. The success of GR as 
the best candidate theory for gravity stems from the fact that it has survived from
most observational tests and is the simplest model among all metric-based geometric 
theories of gravity. However, GR is not without its own problems. To name a few, 
the unavoidable development of singularity signifies the failure of GR at late times, 
the non-renormalizability in the canonical quantization of GR indicates its failure in the early
ages of the universe, and even at the present stage of the universe, GR fails to explain
the accelerated expansion of the universe without resorting to the yet feature-unknown 
dark energy. Some of the problems of GR may be overcame by introducing higher curvature 
terms in the action. For instance, the renormalizability can be significantly improved 
and the accelerated expansion of the universe may be explained without introducing the concept of
dark energy.

A great number of higher curvature gravity models have been proposed
in the literature, each inherits one feature 
of GR or another. To pick out a good replacement of GR among these models, several criteria 
must be considered. First, the alternative model must also pass the observational tests that GR
has survived from. Second, it must resolve at least some of the problems that GR has 
been confronted to. Last but not the least, the alternative model must not introduce 
novel problems which GR did not have. 

The observational tests of GR may be subdivided into two categories. The first is 
consisted of tests on the kinematic level. Most of the well known observational tests 
belong to this category, including, but not limited to, 
the deflection of light, perihelion recession, the time delay of radar signals and the frame 
dragging effect, etc. These are actually tests against the metric 
(Schwarzschild metric in the case of light deflection, perihelion recession and time delay of 
radar signals, Kerr metric in the case of frame dragging effect), rather than against the 
gravitational model, and it is a simple fact that Schwarzschild and Kerr metrics are among 
the universal solutions to all metric-based theories of gravity. The second category involves 
dynamical effects. Only a few observational tests were performed on this level. The direct
observation of gravitational wave and the accelerated expansion of the universe are among this 
category. However, gravitational waves have only been observed in the far field, which 
should be considered as a test of the weak field limit only, and it has been mentioned 
above that GR fails to explain the accelerated expansion of the universe without introducing 
dark energy.  

Among various higher curvature modifications of GR, the Lanczos-Lovelock 
gravity \cite{love1,love2} and $f(R)$ gravity \cite{f(R)1,f(R)2,f(R)3,f(R)4} 
(see also \cite{Nojiri:2010wj,Nojiri:2006ri,Nojiri:2017ncd} for some further 
reviews with cosmological applications) are probably the 
most well-known and intensively studied models. The Lanczos-Lovelock 
gravity modifies GR in such a way that the action is supplemented by a series of 
higher order topological densities, so that in 4-dimensions it falls back to GR
without any modification. $f(R)$ gravity, on the other hand, inherits the fact that the 
Lagrangian density of GR is consisted purely of Ricci scalar, among which GR is the simplest 
choice. We can look at GR from a different angle, i.e. 
its Lagrangian density contains only the metric and the
Ricci tensor. From this point of view, it is natural to extend GR into a more general 
$f$(Ricci) gravity, of which the Ricci polynomial gravity \cite{ricci} proposed recently by us
is a special example. It is also worth mentioning that there has been some research on 
the most general $f$(Riemann) gravity \cite{Tekin:2016vli}, of which all higher curvature 
gravities may be regarded as special cases. However, since the action of $f$(Riemann) 
gravity is too much generic, and the degeneration to special cases is not at all straightforward,
it is still meaningful to study some of its special cases.

In this paper, we will extend the study of Ricci polynomial gravity to a more general 
$f$(Ricci) gravity, where $f(x)$ is an analytic function subjecting to only a few mild 
constraints which guarantees the perturbative behaviors around 
certain Einstein vacua. To be more concrete, we will analyze the perturbative spectrum 
of $f$(Ricci) gravity, identifying the ghost free conditions around all permissible 
Einstein vacua. We also calculate the AdS black hole entropy using the boundary Noether charge 
technique proposed by Majhi and Padmanabhan \cite{mp1,mp2} which is referred to as the MP approach
(see also \cite{jun,Meng} for the use of 
of the same method in the case of $f(R)$ and conformal gravity), because black hole entropy plays
an indespensible role in understanding the holographic properties of the black hole spacetime 
and in exploring the microscopic degrees of freedom of the 
black hole themselves \cite{s1,s2,brow,card,carl,guica}. Our calculation shows 
that the black hole entropy arising from the holographic calculations agrees with Wald's 
geometric entropy.  

The paper is organized as follows. In Sec.2, we introduce the basics of the 
$f$(Ricci) gravity. The bulk and boundary actions, 
the equation of motion and the conditions under which an Einstein manifold can be a 
vacuum solution will be given explicitly. Sec.3 is devoted to the perturbative analysis of 
the model, with emphasis on the ghost free conditions around various maximally symmetric
Einstein vacua. Sec.4 goes beyond the perturbative regime and will be 
concentrated in the calculation of the entropy of spherically symmetric 
black holes using Wald's method and MP approach. 
The paper is then concluded in Sec.5.

\section{The model: equation of motion and vacuum solutions}

To begin with, let us write down the bulk action of the $f$(Ricci) gravity in $n$ dimensions:
\begin{align}
I_{bulk}=\frac{1}{16\pi G}\int_{\mathcal{M}}\,\rd^nx\sqrt{g} \,f(R_{\mu\nu},g_{\mu\nu})
\equiv \frac{1}{16\pi G}\int_{\mathcal{M}}\,\rd^nx\sqrt{g}\,g^{\mu\nu}f_{\mu\nu},
\end{align}
where $f_{\mu\nu} = g_{\mu\rho} f^\rho{}_\nu$, $f^\mu{}_\nu = f(x)|_{x\to R^\mu{}_\nu}$ is 
the natural continuation of $f(x)$, an analytic function in the single 
real variable $x$, to the case of matrix-valued variable, and, of course, $g_{\mu\nu}$ is 
the metric of the spacetime $\mathcal{M}$ with Ricci curvature $R_{\mu\nu}$, and $g$ represents 
the absolute value of $\text{det}(g_{\mu\nu})$. 
When $f(x)$ is a polynomial function, the above action reduces to the Ricci polynomial gravity
studied in \cite{ricci}. Given a coordinate system and assuming that, at certain event in the 
spacetime, all components of the Ricci tensor are small enough, we can understand the 
tensorial expression $f^\mu{}_\nu$ in terms of the Taylor expansion of $f(x)$, i.e.
\begin{align}
f^{\mu}{}_\nu = \sum_{k=0}^\infty \frac{1}{k!} f^{(k)}(0) (\mathcal{R}^{(k)})^\mu{}_\nu, 
\label{expans}
\end{align}
where 
\[
(\mathcal{R}^{(k)})^\mu{}_\nu\equiv R^{\mu}{}_{\mu_1}R^{\mu_1}{}_{\mu_2}\cdots 
R^{\mu_{k-1}}{}_{\nu}.
\]
It is evident that the tensorial monomials $(\mathcal{R}^{(k)})^\mu{}_\nu$ 
are not all independent from expressions of the form $(\mathcal{R}^{(j_1)})^\mu{}_{\mu_1}
(\mathcal{R}^{(j_2)})^{\mu_1}{}_{\mu_2} \cdots
(\mathcal{R}^{(j_p)})^{\mu_{p-1}}{}_\nu R^{k-j_1-\cdots -j_p}$ 
(where $0<j_a<k$ for $a=1,\cdots p$ and $j_1+\cdots+j_p<k$) when $k$ is large enough, 
because the Schouten identity
\[
\delta^{\mu_1\cdots \mu_k}_{\,\nu_1\cdots\nu_k} R^{\nu_1}{}_{\mu_1}\cdots 
R^{\nu_k}{}_{\mu_k}=0
\]
holds identically for $k>n$. This implies that the expanded form \eqref{expans} of $f^{\mu}{}_{\nu}$ 
can be presented in other forms. However, to keep the simplicity of the action, we prefer that 
each term in the expanded form of $f\equiv f(R_{\mu\nu},g_{\mu\nu})=f^{\mu}{}_\mu$ consists in 
a single trace rather than product of multiple traces. This consideration makes the analysis of 
our model much simpler than the cases studied in \cite{1005.1646,1112.6346} and in \cite{Li:2017ncu}. 
Notice also that the works \cite{1005.1646,1112.6346} studied only the cases in three 
dimensions. In generic dimensions, $f$(Ricci) gravity has been considered in 
\cite{Mozaffar:2016hmg}, but the subject there is mainly frame-mapping and 
the calculation of holographic entanglement entropy, which does not overlap with
what we do in this paper.

The process for calculating the first variation of the action is a little bit involved, 
however, the result can be arranged in the following simple form:
\begin{align}
\delta I_{bulk} 
&=\frac{1}{16\pi G}\int_{\mathcal{M}}\, \rd^{n}x\sqrt{g}\,(H_{\mu\nu}\delta g^{\mu\nu}
+\nabla_{\mu}\mathcal{B}^{\mu})\nonumber\\
&= \frac{1}{16\pi G}\left(\int_{\mathcal{M}}\, \rd^{n}x\sqrt{g}\,H_{\mu\nu}\delta g^{\mu\nu}
+\int_{\partial \mathcal{M}}\, \rd^{n-1}x\sqrt{g_{(n-1)}}\,n_\mu \mathcal{B}^{\mu}\right)
, \label{var}
\end{align}
where
\begin{align}
H_{\mu\nu}&\equiv -\frac{1}{2}fg_{\mu\nu}+f'{}_{(\mu}{}^{\sigma} R_{\nu)\sigma}
+\frac{1}{2}\Box f'{}_{\mu\nu}-\nabla_{\rho}\nabla_{(\mu}f'{}_{\nu)}{}^{\rho}
+\frac{1}{2}\nabla_{\rho}\nabla_{\sigma}f'{}^{\rho\sigma}g_{\mu\nu}, \label{hmunu}\\
\mathcal{B}^{\mu}
&=f'{}^{\rho\nu}\delta\Gamma^{\mu}_{\rho\nu}-f'{}^{\rho\mu}\delta\Gamma^{\nu}_{\nu\rho}
+\nabla_{\rho}f'{}^{\mu}{}_{\sigma}\delta g^{\rho\sigma}
-\frac{1}{2}\nabla^{\mu}f'{}_{\rho\sigma}\delta g^{\rho\sigma}
-\frac{1}{2}g_{\rho\sigma}\nabla_{\nu}f'{}^{\mu\nu}\delta g^{\rho\sigma}, \label{bmu}
\end{align}
$g_{(n-1)}$ is the absolute value of the determinant of the induced metric on the spacetime 
boundary $\partial \mathcal{M}$, $n_\mu$ is the unit outer-pointing normal covector of 
$\partial \mathcal{M}$, and $f'{}^\mu{}_\nu$ is defined in the same spirit of $f^\mu{}_\nu$, 
with $f(x)$ replaced by $f'(x)$. In eq.\eqref{var}, the term involving 
$n_{\mu}\mathcal{B}^{\mu}$ cannot be simply dropped by imposing the fixed boundary condition 
$\delta g_{\mu\nu}|_{\partial\mathcal{M}}=0$, because the first two terms on the right hand side of 
eq.\eqref{bmu} contains derivatives of $\delta g_{\mu\nu}$ rather than $\delta g_{\mu\nu}$ itself.
To get rid of this redundant term, some boundary action must be introduced. 
One way to achieve this is to adopt the method introduced in \cite{zhb}, which works for any 
higher curvature gravity with Lagrangian density composed of the Riemann tensor and the metric.
Schematically, one introduces two auxiliary fields $\phi_{\mu\nu\rho\sigma}$ and 
$\psi_{\mu\nu\rho\sigma}$ to rearrange the bulk action into a form 
which is linear in the Riemann tensor,
\begin{equation}
I_{bulk}=\int_{\mathcal{M}}\, \rd^{n}x\sqrt{g}\,
\left[ f(\phi_{\mu\nu\rho\sigma},g_{\mu\nu})
-\psi^{\mu\nu\rho\sigma}(\phi_{\mu\nu\rho\sigma}-R_{\mu\nu\rho\sigma}) \right], \label{lin1}
\end{equation}
where $f(\phi_{\mu\nu\rho\sigma},g_{\mu\nu})$ is nothing but $f(R_{\mu\nu},g_{\mu\nu})$ with
all occurrences of the Riemann tensor $R_{\mu\nu\rho\sigma}$ replaced by $\phi_{\mu\nu\rho\sigma}$. 
Then, following a systematic variation process one finds that the only source of necessity for a 
boundary action comes from the variation of $R_{\mu\nu\rho\sigma}$. Finally, the boundary action
can be written as
\begin{align}
I_{bdry}&=\frac{1}{16\pi G}\int_{\partial\mathcal{M}}\, \rd^{n-1}x\sqrt{g_{(n-1)}}
\,\mathcal{L}_{B},  \label{ba}
\end{align}
where the boundary Lagrangian density $\mathcal{L}_{B}$ is given by
\begin{align}
\mathcal{L}_{B}&=f'{}^{\mu\nu}(K_{\mu\nu}+n_{\mu}n_{\nu}K), \label{surf1}
\end{align}
wherein $K_{\mu\nu}=\nabla_{\mu}n_{\nu}$ is the extrinsic curvature tensor and $K$ is 
the trace of $K_{\mu\nu}$. 

It is a common knowledge that the choice of the boundary action is not unique: two different 
boundary actions that differ from each other only by some terms vanishing on the boundary 
will be equally good to make the variational problem self consistent. In our case, 
instead of eq.\eqref{lin1}, we can make the bulk action linear in the Ricci tensor rather 
than in the Riemann tensor by writing
\begin{align}
I_{bulk}=\int_{\mathcal{M}}\, \rd^{n}x\sqrt{g}\,
\left[ f(\phi_{\mu\nu},g_{\mu\nu})
-\psi^{\mu\nu}(\phi_{\mu\nu}-R_{\mu\nu}) \right]. \label{lin2}
\end{align}
Then, a similar procedure yields the boundary action \eqref{ba} with $\mathcal{L}_B$ replaced by
\begin{align}
\mathcal{L}_{B}&= n_{\mu} \Big(f'{}^{\rho\mu}\Gamma^{\nu}_{\nu\rho}
-f'{}^{\rho\nu}\Gamma^{\mu}_{\rho\nu}
\Big). \label{surf2}
\end{align}

With the aid of either eq.\eqref{surf1} or eq.\eqref{surf2}, we can make the variational problem
of our model consistent and hence it follows that the equation of motion of the model is simply
\begin{align}
H_{\mu\nu}=0.  \label{eom}
\end{align}
Recalling the concrete form \eqref{hmunu} of $H_{\mu\nu}$ and assuming that the 
equation of motion admits an Einstein metric obeying
\begin{align}
R_{\mu\nu}=\chi g_{\mu\nu} \label{einm}
\end{align}
as an exact solution, where the constant $\chi$ is related to the cosmological constant $\Lambda$ 
via $\chi=\frac{2\Lambda}{n-2}$, it follows from eq.\eqref{eom} that $\chi$ has to solve 
the following algebraic equation:
\begin{align}
\chi f'(\chi)-\frac{n}{2}f(\chi) =0.  \label{suffness}
\end{align}
Let us stress that this is an algebraic equation for $\chi$, not a differential equation for 
$f(\chi)$, because $f(x)$ is prescribed while defining the model. 
However, if one happens to have chosen the function $f(x)$ to be proportional to $x^{n/2}$
then the above equation will be identically satisfied for any value of $\chi$.  
In the particular case of $n=4$, this amounts to $f(x)\sim x^2$, the corresponding
model is simply the Ricci squared gravity with bulk Lagrangian density 
$\mathcal{L}\propto \mathcal{R}^{(2)}$. For generic choice of $f(x)$, eq.\eqref{suffness} is
the necessary and sufficient condition in order that the Einstein manifolds \eqref{einm} can 
arise as exact solutions to our model, and the permissible values for $\chi$ may not be 
unique. This fact has already become evident in the earlier study \cite{ricci} about the 
Ricci polynomial gravity.

\section{Perturbative properties around Einstein vacua}

In this section we will analyze the perturbative properties of the $f$(Ricci) gravity 
around the background Einstein metric $g_{\mu\nu}$ satisfying the equation
$R_{\mu\nu}(g)=\chi g_{\mu\nu}$. The metric with fluctuation may be denoted by $\tilde{g}_{\mu\nu}
=g_{\mu\nu}+h_{\mu\nu}$, where $h_{\mu\nu}$ is a small deviation from the background metric. 
It is customary to denote $\nabla_{\nu}h^{\mu\nu}=A_{\mu}$, where $\nabla_{\mu}$ is the 
covariant derivative compatible with $g_{\mu\nu}$. We also denote 
the traceless part of $h_{\mu\nu}$ by 
$\bar{h}_{\mu\nu}=h_{\mu\nu}-\frac{1}{n}hg_{\mu\nu}$, where $h=g^{\mu\nu}h_{\mu\nu}$.

Up to linear level in $h_{\mu\nu}$, the perturbed equation of motion reads
\begin{align}
\delta H_{\mu\nu}
=&-\left[\frac{1}{2}nf(\chi) +\chi^2f''(\chi)\right] h_{\mu\nu}
  +\frac{1}{2}g_{\mu\nu} f'(\chi)\left[\chi h
  +\frac{1}{2}g^{\rho\sigma}\Delta_{L}h_{\rho\sigma}\right]   \nonumber  \\
&-\frac{1}{2}\left[ \chi  f''(\chi)+f'(\chi)\right]\Delta_{L}h_{\mu\nu}
+ f''(\chi)\left[\frac{1}{2}\nabla^{\rho}\nabla_{(\mu}\Delta_{L}h_{\nu)\rho}
+\chi \nabla^{\rho}\nabla_{(\mu}h_{\nu)\rho}\right]  \nonumber   \\
&-\frac{1}{2}f''(\chi)g_{\mu\nu}\left[\frac{1}{2}\nabla^{\rho}\nabla^{\sigma}\Delta_{L}h_{\rho\sigma}
+\chi \nabla^{\rho}\nabla^{\sigma}h_{\rho\sigma}\right] \nonumber   \\
&-\frac{1}{2}f''(\chi)\left[\frac{1}{2}\Box \Delta_{L}h_{\mu\nu}+\chi \Box h_{\mu\nu}\right]\nonumber\\
=&0,   \label{genlin}
\end{align}
where
\begin{align}
\Delta_{L}h_{\mu\nu}= \Box h_{\mu\nu}+\nabla_{\mu}\nabla_{\nu}h
-2[\nabla_{(\mu} A_{\nu)}+R^{\rho}{}_{(\mu} h_{\nu)\rho}
-R_{\mu}{}^\rho{}_\nu{}^\sigma h_{\rho\sigma}].
\end{align}
The operator $\Delta_{L}$ is known as the Lichnerowicz operator.

We will be particularly interested in the perturbative behavior around the maximally symmetric 
vacua, because the linearized equation of motion can be further simplified 
using the following properties which any maximally symmetric vacuum must obey:
\begin{align}
R_{\mu\nu\rho\sigma}=
\frac{\chi}{n-1}[g_{\mu\rho}g_{\nu\sigma}-g_{\mu\sigma}g_{\nu\rho}], \quad 
R_{\mu\nu}=\chi g_{\mu\nu},\quad 
R=n\chi.
\end{align}
In such backgrounds, the action of the Lichnerowicz operator on $h_{\mu\nu}$ is
simplified into
\begin{align*}
\Delta_{L}h_{\mu\nu}= \Box h_{\mu\nu}+\nabla_{\mu}\nabla_{\nu}h
-2\nabla_{(\mu} A_{\nu)}+\frac{2\chi}{n-1} g_{\mu\nu}h
-\frac{2n\chi}{n-1} h_{\mu\nu}.	
\end{align*}
In the following, two sub-cases with $\chi=0$ and $\chi\not=0$ will be considered seperately.

\subsection{Minkowski background with $\chi=0$}

The Minkowski spacetime corresponds to the choice $\chi=0$. According to eq.\eqref{suffness}, 
an Einstein manifold with $\chi=0$ is an exact vacuum solution of $f$(Ricci) gravity 
if and only if $f(0)=0$. Therefore, for the Minkowski background, the linearized equation of 
motion can be greatly simplified:
\begin{align}
\delta H_{\mu\nu}
=&-\frac{1}{4}f''(0)\left[\Box^2 h_{\mu\nu}
+\eta_{\mu\nu}(\Box^2h-\Box\partial^{\rho}A_{\rho}) 
-\Box\partial_{\mu}\partial_{\nu}h
-2\Box\partial_{(\mu}A_{\nu)}
+2\partial_{(\mu}\partial_{\nu)}\partial^{\rho}A_{\rho})\right]   \nonumber   \\
&-\frac{1}{2}f'(0)\left[\Box h_{\mu\nu}-\eta_{\mu\nu}(\Box h-\partial^{\rho}A_{\rho})+\partial_{\mu}\partial_{\nu}h
-2\partial_{(\mu}A_{\nu)}\right].   
\end{align}

To further analyze the perturbative spectrum of the model, we consider separately the 
two different cases $f''(0)=0$ and $f''(0)\neq0$. When $f''(0)=0$, the linearized equation 
of motion will beome the same as the linearized standard Einstein equation (we assume 
that $f'(0)\neq 0$ in this case), which, after choosing the transverse traceless gauge, describes 
a massless spin-2 field and so the fluctuation around the Minkowski vacuum is ghost-free.
On the other hand, if we set $f''(0)\neq 0$ and choose the gauge fixing condition 
$A_{\mu}=\frac{1}{n}\partial_{\mu}h$, the linearized equation of motion becomes
\begin{align}
\delta H_{\mu\nu}
&=-\frac{1}{4}f''(0)
\left[\Box (\Box- m_1^2)\bar{h}_{\mu\nu}
+(\eta_{\mu\nu}\Box-\partial_{\mu}\partial_{\nu})\left(\Box-m_2^2 \right)h\right],
\end{align}
where 
\[
m_1^2=-\frac{2f'(0)}{f''(0)},\quad
m_2^2=\frac{2(n-2)}{n}\frac{f'(0)}{f''(0)}.
\]  
In this case, if $f'(0)=0$, we have $m_1^2=m_2^2=0$ and hence $\bar{h}_{\mu\nu}$ and $h$ 
respectively correspond to a massless spin-2 and a massless spin-0 mode. If, however, 
$f'(0)\neq0$, then either $m_1^2$ or $m_2^2$ will be negative, and the corresponding mode
is exactly what is refered to as ghost mode.

Thus we conclude that for Minkowski background, the ghost-free condition is either 
$f''(0)=0$ with $f'(0)\neq0$ or $f''(0)\neq 0$ with $f'(0)= 0$. Notice that, in the special case 
of Ricci polynomial gravity \cite{ricci}, we have fixed the coefficient in front of the 
first order term in the Lagrangian density to be unity, which implies $f'(0)=1$, therefore the 
latter ghost free condition $f''(0)\neq 0$ with $f'(0)= 0$ was not seen in \cite{ricci}.

\subsection{(A)dS background}

The maximally symmetric background with $\chi\neq 0$ is either de Sitter (dS) or 
anti-de Sitter (AdS). In this subsection we will consider the perturbative spectrum around 
these two types of backgrounds simultaneously.

In (A)dS background, we can choose the gauge fixing condition $A_{\mu}=\nabla_{\mu}h$ and 
making use of eq.\eqref{suffness} to change all occurrences of $f'(\chi)$ into a 
multiple of $f(\chi)$, after which eq.\eqref{genlin} becomes
\begin{align}
\delta H_{\mu\nu}
=&-\frac{1}{4}f''(\chi)\left[\Box^2h_{\mu\nu}-\nabla_{(\mu}\nabla_{\nu)}\Box h\right]
+\left[\frac{n}{2(n-1)}f(\chi)-\frac{1}{(n-1)^2}\chi^2f''(\chi)\right]h_{\mu\nu} 
\nonumber \\
&+\left[-\frac{n}{4}\frac{f(\chi)}{\chi}
+\frac{1}{n-1}\chi f''(\chi)\right]\Box h_{\mu\nu} 
-\frac{n+3}{4(n-1)}\chi f''(\chi)g_{\mu\nu}\Box h  \nonumber  \\
&+\left[\frac{n}{4}\frac{f(\chi)}{\chi}+\chi f''(\chi)\right]\nabla_{\mu}\nabla_{\nu}h
+\left[\frac{n(n-3)}{4(n-1)}f(\chi)-\frac{n-2}{(n-1)^2}\chi^2f''(\chi)\right]g_{\mu\nu}h
\nonumber\\
=&0.
\label{ads}  
\end{align}

We will again proceed by considering the two choices $f''(\chi)=0$ and $f''(\chi)\neq0$
separately.

When $f''(\chi)=0$, eq.\eqref{ads} can be simplified into
\begin{align}
\frac{n}{2(n-1)}f(\chi)h_{\mu\nu}
-\frac{n}{4}\frac{f(\chi)}{\chi}\Box h_{\mu\nu}
+\frac{n}{4}\frac{f(\chi)}{\chi}\nabla_{\mu}\nabla_{\nu}h
+\frac{n(n-3)}{4(n-1)}f(\chi)hg_{\mu\nu}=0.
\end{align}
Taking the trace by contraction with $g^{\mu\nu}$, we get
\begin{align}
\frac{n(n-1)}{4}f(\chi)h=0,
\end{align}
i.e. $h=0$. Thus eq.\eqref{ads} becomes
\begin{align}
\delta H_{\mu\nu}
=-\frac{n}{4}\frac{f(\chi)}{\chi}\left[\Box-\frac{2\chi}{n-1}\right]\bar{h}_{\mu\nu}=0,
\end{align}
which implies that $\bar{h}_{\mu\nu}$ a massless spin-2 mode in (A)dS background. 

When $f''(\chi)\neq0$, we can separate eq.\eqref{ads} into the traceless and trace parts,
i.e.
\begin{align}
\delta H_{\mu\nu}=\delta H_{\mu\nu}^{(1)}+\delta H_{\mu\nu}^{(2)}=0,
\label{adspert}
\end{align}
where
\begin{align}
\delta H_{\mu\nu}^{(1)}=&-\frac{1}{4}f''(\chi)\left(\Box-\frac{2\chi}{n-1}-m^2_1\right)
\left(\Box-\frac{2\chi}{n-1}-m^2_2\right)\bar{h}_{\mu\nu},	
\end{align}
with
\begin{align}
m^2_1=0,\quad m^2_2=-n\frac{f(\chi)}{\chi f''(\chi)},
\end{align}
and
\begin{align}
\delta H_{\mu\nu}^{(2)}
=&-\frac{1}{4}f''(\chi)\left[\frac{1}{n}g_{\mu\nu}\Box - \nabla_{(\mu}\nabla_{\nu)}\right] 
\left[\Box + \left(\frac{n f(\chi)}{\chi f''(\chi)}+4\chi \right) \right]h
\nonumber\\
&-\frac{1}{4}\chi f''(\chi) g_{\mu\nu}\left\{\Box   
-\left[ \frac{(n-2) f(\chi)}{\chi f''(\chi)}-\frac{4\chi}{n} \right]\right\}h.
\label{dH2}
\end{align}
Taking the trace of the above equation, we get
\begin{align}
\delta H^{(2)}= -\frac{n}{4}\chi f''(\chi)\Big(\Box-m^2_3\Big)h=0,   \label{scalar-0}
\end{align}
where
\begin{equation}
m^2_3=\frac{(n-2) f(\chi)}{\chi f''(\chi)}-\frac{4\chi}{n} . \label{cc}
\end{equation}

Now because of $\bar h_{\mu\nu}$ and $h$ are essentially independent, eq.\eqref{adspert}
implies both $\delta H_{\mu\nu}^{(1)}=0$ and $\delta H_{\mu\nu}^{(2)}=0$. It follows from the 
first of these two conditions that $\bar h_{\mu\nu}$ is consisted of two traceless spin-2 modes,
one of which is massless and the other is either a massive or a ghost mode. The condition that 
$\bar h_{\mu\nu}$ does not contain a ghost mode is $m_2^2\geq 0$, or explicitly
\begin{align}
\frac{f(\chi)}{\chi f''(\chi)}\leq 0. \label{chimin0}
\end{align}
The second condition $\delta H_{\mu\nu}^{(2)}=0$ looks more complicated. According to 
eq.\eqref{scalar-0}, the last term in eq.\eqref{dH2} can be dropped. Then the remaining terms
in eq.\eqref{dH2} implies that $h$ subjects to some extra constraints in addition to 
the wave-like equation \eqref{scalar-0}, unless the following condition is satisfied,
\begin{equation}
m^2_3=\frac{(n-2) f(\chi)}{\chi f''(\chi)}-\frac{4\chi}{n} 
= -\left(\frac{n f(\chi)}{\chi f''(\chi)}+4\chi \right),
\label{ccc}
\end{equation}
in which case eq.\eqref{dH2} can be rewritten in a completely factorized form, i.e.
\begin{align}
\delta H_{\mu\nu}^{(2)}
=&-\frac{1}{4}f''(\chi)\left[\frac{1}{n}g_{\mu\nu}\Box - \nabla_{(\mu}\nabla_{\nu)}
+\chi g_{\mu\nu}\right] \left(\Box - m_3^2 \right)h.
\end{align}

Eq.\eqref{ccc} can be simplified into
\begin{align}
\frac{f(\chi}{f''(\chi)}= - \frac{2\chi^2}{n}.	\label{32}
\end{align}
Now, since the condition that $h$ does not correspond
to a scalar ghost is
\[
m^2_3=\frac{(n-2) f(\chi)}{\chi f''(\chi)}-\frac{4\chi}{n} \geq 0,
\]
which gives
\begin{align}
\chi \leq \frac{n(n-2)}{4}\frac{f(\chi)}{\chi f''(\chi)},  \label{chimin}
\end{align}
we have from eq.\eqref{chimin0} that $\chi<0$ and 
$\frac{f(\chi)}{f''(\chi)} > 0$.
However, eq.\eqref{32} requires $\frac{f(\chi)}{f''(\chi)}<0$. Therefore we conclude that
for $f''(\chi)\neq 0$, the model cannot be ghost free around the maximally symmetric
Einstein vacua with $\chi\neq 0$. In other words, $f$(Ricci) gravity can only 
be ghost free around the maximally symmetric Einstein vacua when 
$f''(\chi)=0$. Let us remind that, the condition $f''(\chi)=0$ is identical to 
the ghost free condition presented in \cite{ricci} for the special case of 
Ricci polynomial gravity.

\section{Black hole entropy}

It is clear from Sec.2 that, provided the condition \eqref{suffness} is satisfied, 
an Einstein manifold obeying eq.\eqref{einm} will be a vacuum solution 
of our model. In this section, we are particularly interested in the spherically 
symmetric black hole solutions and will be concentrated in the calculation of black hole entropies
for such solutions. Unlike the standard GR, the commonly acknowledged form for the black 
hole entropy associated with higher curvature gravities is not the Bekenstein-Hawking entropy,
but rather Wald's geometric entropy. We will show that the Wald entropy is identical to 
the holographic entropy calculated using the MP approach.

For any higher curvature gravity with bulk action
\[
I_{bulk}=\int \rd^n x\sqrt{g} \,L(g_{\mu\nu},R_{\mu\nu\rho\sigma}),
\]
the Wald entropy associated with a spherically symmetric black hole solution
of the form (here we take a coordinate system $x^\mu=(x^0,x^1,\cdots x^{n-1})
=(t,r, x^2,\cdots x^{n-1})$)
\begin{align}
\rd s^2=-\tilde{f}(r)\rd t^2+\frac{1}{\tilde{f}(r)}\rd r^2+r^2\Omega_{ij}\rd x^{i}\rd x^{j},
\quad(i,j=2,\cdots, n-1)
\label{bh}
\end{align}
can be evaluated via the following formula:
\begin{align}
S_{Wald}=-2\pi \int \rd^{n-2}x\sqrt{g_{(n-2)}}
\frac{\delta L}{\delta R_{\alpha\beta\gamma\delta}}
\epsilon_{\alpha\beta}\epsilon_{\gamma\delta}, \label{Wald1}
\end{align}
where $\Omega_{ij}$ is the metric on a $(n-2)$-dimensional unit sphere and $
g_{(n-2)}=|\mbox{det}(r_h^2 \Omega_{ij})|$, with $r_h$ being the radius of the black hole 
event horizon, and $\epsilon_{\alpha\beta}$ is given via $\epsilon_{01}=-\epsilon_{10}=1$.
Note that the integration in the above formula is taken over the compact spacial section of 
the horizon hypersurface, a sphere with radius $r_h$.

In our case, $L=\frac{1}{16\pi G} f(R_{\mu\nu},g_{\mu\nu})$. Therefore, the Wald entropy 
\eqref{Wald1} becomes
\begin{align}
S_{Wald}=&-\frac{1}{8 G} \int \rd^{n-2}x \sqrt{g_{(n-2)}}
\frac{\delta f}{\delta R_{\alpha\beta\gamma\delta}}
\epsilon_{\alpha\beta}\epsilon_{\gamma\delta} \nonumber\\
=&-\frac{1}{8 G} \int \rd^{n-2}x \sqrt{g_{(n-2)}} f'^{\rho\sigma}g^{\xi\eta}
\frac{\delta R_{\rho\xi\sigma\eta}}{\delta R_{\alpha\beta\gamma\delta}}
\epsilon_{\alpha\beta}\epsilon_{\gamma\delta}.
\end{align}
Using the identity \cite{Miao}
\begin{align*}
\frac{\delta R_{a_1b_1c_1d_1}}{\delta R_{abcd}}
&=\frac{1}{12}\bigg(
\delta^{ab}_{a_1b_1}\delta^{cd}_{c_1d_1}
-\frac{1}{2}\delta^{ac}_{a_1b_1}\delta^{db}_{c_1d_1}
-\frac{1}{2}\delta^{ad}_{a_1b_1}\delta^{bc}_{c_1d_1}  \\
&\quad +\delta^{cd}_{a_1b_1}\delta^{ab}_{c_1d_1}
-\frac{1}{2}\delta^{db}_{a_1b_1}\delta^{ac}_{c_1d_1}
-\frac{1}{2}\delta^{bc}_{a_1b_1}\delta^{ad}_{c_1d_1}\bigg)
\end{align*}
and the explicit form of $\epsilon_{\alpha\beta}$, we can reduce the expression for the 
Wald entropy into the final result
\begin{align}
S_{Wald}
=&\frac{A}{8G}\Big(f'{}^{0}{}_{0}+f'{}^{1}{}_{1}\Big)_{r\to r_h},  \label{walden}
\end{align}
where 
\begin{align}
A=\int \rd^{n-2}x \sqrt{g_{(n-2)}}|_{r=r_h}  \label{area}
\end{align}
is the area of the event horizon.
It is a trivial practice to verify that for the standard GR, the above result reduces into the
Bekenstein-Hawking entropy $S=\frac{A}{4G}$.

Now let us reconsider the black hole entropy from a holographic point of view. 
There are several methods to pursue the holographic calculation of black hole entropy.
The earliest attempt in this direction was made by Brown and Henneaux \cite{brow}, who
successfully calculated the asymptotic Virasoro symmetry for three dimensional AdS spacetime
by assuming some mild boundary conditions. Their work was further developed by 
Strominger \cite{s1,s2} and Carlip \cite{carl,Carlip2} respectively to the case of black 
hole spacetimes, and, by use of the Cardy formula \cite{card}, they were able to calculate 
the entropy of various black holes. In this approach, 
the bulk action must be made use of. Alternatively, the MP approach makes use of only the 
boundary action and the process for obtaining the black hole entropy is much simpler.
Therefore we shall adopt the MP approach in the following calculation. 

The MP approach is applicable to any metric-based geometric theory of gravity. The key 
ingredient in this approach is the boundary Noether charge associated with the 
asymptotic diffeomorphism $x^\mu\to x^\mu+\xi^\mu(x)$, where the spacetime boundary 
$\partial\mathcal{M}$ is 
taken to be a near-horizon hypersurface. For generic boundary action of 
the form \eqref{ba}, the boundary Noether charge can be evaluated via
\begin{align}
Q[\xi]=\frac{1}{2}\int \sqrt{h} \,\rd\Sigma_{\mu\nu} J^{\mu\nu},
\end{align}
where
\[
\rd\Sigma_{\mu\nu} = \rd^{n-2}x (n_\mu m_\nu-m_\mu n_\nu)
\]
is the area element on the constant-time slice $\Sigma$ of the near-horizon hypersurface
$\partial\mathcal{M}$, $n_\mu$ and $m_\mu$ are respectively the unit outer-pointing spacelike 
normal covevtor and the unit future pointing timelike normal covector of $\Sigma$, 
$h$ is the determinant of the induced metric on $\Sigma$ (which reduces into $g_{(n-2)}$ 
described above in the near-horizon limit), and 
\begin{align}
J^{\mu\nu}[\xi]=\frac{1}{16\pi G}\mathcal{L}_{B}[\xi^{\mu}n^{\nu}-\xi^{\nu}n^{\mu}]
\end{align}
is known as the Noether potential \cite{mp1,mp2,Meng,jun}. Notice that $n_\mu$ is identical to 
the unit normal covector that appeared in the boundary action.
In the following, we shall take the boundary 
Lagrangian density \eqref{surf1} as working example and leave for the readers to check that 
the alternative boundary Lagrangian density \eqref{surf2} works equally well.

Using the local Rindler coordinate $\rho=r-r_h$, the metric \eqref{bh} can be rewritten as
\begin{align}
\rd s^2=-\tilde{f}(r_h+\rho)\rd t^2+\frac{1}{\tilde{f}(r_h+\rho)}\rd \rho^2
+r^2\Omega_{ij}\rd x^{i}\rd x^{j}.
\end{align}
In this spacetime, the unit normal vectors of the black hole event horizon can be chosen as
\begin{align}
n^{\mu}=\left(0,\sqrt{\tilde{f}(r)},0,\cdots,0\right),\quad
m^{\mu}=\left(\frac{1}{\sqrt{\tilde{f}(r)}},0,0,\cdots,0\right),
\end{align}
and the generator $\xi^{\mu}$ of the boundary diffeomorphism is
\begin{align}
\xi^{t}=T-\frac{\rho}{\tilde{f}(r_H+\rho)}\partial_tT,\ \ 
\xi^{\rho}=-\rho\partial_tT,
\end{align}
where $T=T(t,\rho)$ is an arbitrary function. We can expand $T$ in term of a set of 
basis functions
\begin{align}
T=\sum a_mT_m,
\end{align}
where the basis functions $T_m$ should satisfy the Diff$(S^1)$ algebra. A standard choice is
\begin{align}
T_m=\frac{1}{\alpha}\exp{[\mbox{i}m(\alpha t+g(\rho)+p\cdot x)]},
\end{align}
where $\alpha$ is a constant, $p$ is an integer and $g(\rho)$ is a regular function on 
the horizon. Clearly, we must have $a^{*}_m=a_{-m}$ in order to make $T$ real.

With the above prescription, the boundary Noether charge can be expressed as
\begin{align}
Q[\xi]=\frac{1}{16\pi G}\int\rd^{n-2}x\sqrt{g_{(n-2)}}\left[f'{}^{t}{}_{t}
+f'{}^{\rho}{}_{\rho}\right]\big|_{\rho\to 0}\left(\kappa T-\frac{1}{2}\partial_t T\right)
\end{align}
in the near horizon limit $\rho\to 0$. 
The commutator of between two charges is then given by
\begin{align}
\left[Q[\xi_1],Q[\xi_2]\right]
&=\frac{1}{16\pi G}\int\rd^{n-2}x\sqrt{g_{(n-2)}}
\left[f'{}^{t}{}_{t}+f'{}^{\rho}{}_{\rho}\right]\big|_{\rho\to 0}
\bigg[\kappa \left(T_1\partial_tT_2-T_2\partial_tT_1\right) \nonumber  \\
&-\frac{1}{2}\left(T_1\partial_t^2T_2-T_2\partial_t^2T_1\right)
+\frac{1}{4\kappa}\left(\partial_tT_1\partial_t^2T_2-\partial_tT_2\partial_t^2T_1\right)\bigg],
\end{align}
where $\kappa=\frac{\tilde{f}'(r_h)}{2}$.
Denoting with $Q_m$ the mode corresponding to $T_m$, we have
\begin{align}
Q_m&=\frac{1}{16\pi G}\left[f'{}^{t}{}_{t}+f'{}^{\rho}{}_{\rho}\right]\big|_{\rho\to 0}
\frac{\kappa A}{\alpha}\delta_{m,0},
\\
[Q_m,Q_n]&=-\mbox{i}(m-n)Q_{m+n}-\mbox{i}m^3\frac{C}{12}\delta_{m+n,0},
\label{vir}
\end{align}
where $A$ is given in eq.\eqref{area}, and
\begin{align}
C=\frac{3}{8\pi G}\left[f'{}^{t}{}_{t}+f'{}^{\rho}{}_{\rho}\right]\big|_{\rho\to 0}
\left(\frac{\alpha A}{\kappa}\right). \label{cen}
\end{align}
Notice that the commutator \eqref{vir} is actually the celebrated Virasoro algebra with
the generator $Q_0$ shifted by a constant, wherein the 
central charge $C$ given by eq.\eqref{cen}. 
This observation ensures that the well-known Cardy's
formula is applicable in the present case.  
Finally, using Cardy's formula, the entropy of the black hole is evaluated to be
\begin{align}
S=2\pi\sqrt{\frac{CQ_0}{6}}
=\frac{A}{8G}\left[f'{}^{t}{}_{t}+f'{}^{\rho}{}_{\rho}\right]\big|_{\rho\to 0}.
\label{mpen}
\end{align}
This result is in exact agreement with the Wald entropy \eqref{walden}.

Before closing, it may be intriguing to make a comparison between the black hole entropies
between $f$(Ricci) gravity and $f(R)$ gravity for the same $f(x)$ and the same black hole metric.
To be more concrete, we consider the Tangherlini-(A)dS black hole solution
\begin{align}
\rd s^2=-\left(1-\frac{2M}{r^{n-3}}-\frac{\chi r^2}{n-1}\right)\rd t^2
+\left(1-\frac{2M}{r^{n-3}}-\frac{\chi r^2}{n-1}\right)^{-1}\rd r^2
+r^2\Omega_{ij}\rd x^{i}\rd x^{j},
\end{align}
which obviously obeys \eqref{einm}. For this solution, the corresponding entropy in $f$(Ricci)
gravity, i.e. eq.\eqref{walden} or \eqref{mpen}, reduces into
\begin{align}
S_{f(\text{Ricci})}	= \frac{A}{4G} f'(\chi).
\end{align}
In comparison, the same black hole solution in $f(R)$ gravity has entropy \cite{jun}
\begin{align}
S_{f(R)}= \frac{A}{4G} f'(R) = \frac{A}{4G} f'(n \chi).
\end{align}

\section{Conclusions}

$f$(Ricci) gravity is an extension the Ricci polynomial gravity proposed in \cite{ricci}.
It turns out that $f$(Ricci) gravity has quite some elegant features. 
The model can have multiple Einstein vacua and it can be ghost free around the 
maximally symmetric Einstein vacua under only a few mild constraint. 
Beyond the perturbative regime, we have calculated the 
entropy of spherically symmetric black hole solutions using both Wald's entropy formula
and the MP approach, and both results are in good agreement. Similar analasys was performed
for BTZ black hole solutions for $f$(Ricci) gravity in three-dimensions \cite{SS}.
We expect that $f$(Ricci) gravity may become a competitive gravitational model in the 
future researches.

\section*{Acknowledgement}

This work is supported by the National Natural Science Foundation of China under the grant
No. 11575088. We would like to thank Bin Wu for helpful discussions and also Wenli Yang and Zhanying
Yang for hospitality at the Institute of Modern Physics, Norsthwest University during the 
completion of this work.

\providecommand{\href}[2]{#2}\begingroup
\footnotesize\itemsep=0pt
\providecommand{\eprint}[2][]{\href{http://arxiv.org/abs/#2}{arXiv:#2}}


\end{document}